# Simplified equations for the semi-localized transitions (SLT) model

**Rafał PORZEZIŃSKI, Arkadiusz MANDOWSKI**

*Institute of Physics, Jan Dlugosz University, ul. Armii Krajowej 13/15, PL-42200 Częstochowa, Poland; rafal.porzezinski@doktorant.ujd.edu.pl*

**Abstract**

A simplified version of the semi-localized transitions (SLT) model is derived. The SLT-QE5 approximation reduces the number of differential equations from seven to five. Numerical calculations show that the approximation is very good for various trap parameters except the case with large reduction of activation energy during recombination to adjacent hole together with very low retrapping coefficient.

Highligths:

- New approximation was derived for the model of semi-localized transitions (SLT)
- The SLT-QE5 approximation reduces the number of differential equations from seven to five.
- Numerical calculations show that the approximation is very good for most trap parameters
- The SLT-QE5 approximation accuracy increases with increasing reatrapping

# Simplified equations for the semi-localized transitions (SLT) model

Rafał PORZEZIŃSKI, Arkadiusz MANDOWSKI

*Institute of Physics, Jan Dlugosz University, ul. Armii Krajowej 13/15, PL-42200 Częstochowa, Poland; rafal.porzezinski@doktorant.ujd.edu.pl*

**Abstract**

A simplified version of the semi-localized transitions (SLT) model is derived. The SLT-QE5 approximation reduces the number of differential equations from seven to five. Numerical calculations show that the approximation is very good for various trap parameters except the case with large reduction of activation energy during recombination to adjacent hole together with very low retrapping coefficient.

## 1. Introduction

Thermoluminescence (TL) and Optically Stimulated Luminescence (OSL) are widely used for radiation dosimetry and dating. The kinetics of charge carrier trapping and recombination underlying these processes is complex. While classical models like the Simple Trap Model (STM) (Chen and Mckeever, 1997) and the Localized Transition (LT) model (Halperin and Braner, 1960; Land, 1969) exist, they often fail to fully explain experimental observations. The Semi-Localized Transition (SLT) model, proposed by Mandowski (Mandowski, 2005), offers a more comprehensive framework by incorporating both localized recombination within trap-recombination center (T-RC) pairs and delocalized transitions via transport bands SLT successfully accounts for phenomena like displacement peaks and anomalously high frequency factors (Mandowski, 2006, 2008).

Research on the SLT model, its applications, and modifications is actively being conducted (Pagonis, 2005; Kumar et al., 2007, 2006, 2011). Mandowski and Bos explained an anomalous heating rate effect in thermoluminescence using the SLT model (Mandowski and Bos, 2011). In 2013 Pagonis et al. (Pagonis et al., 2013) demonstrated that this phenomenon can also be described using a simplified SLT model. Additionally, a two-level version of the semi-localized transitions (SLT) model for a thermoluminescence dosimeter was proposed by Sadri et al. (Sadri et al., 2019a, 2019b). Concurrently, Eliyahu, Horowitz, and colleagues (Eliyahu et al., 2014) developed conduction band/valence band kinetic model for thermoluminescence that incorporates both localized and delocalized recombination mechanisms to explain the dependence of TL dose response on photon energy. Their model, similar to SLT, is based on the assumption of the simultaneous occurrence of both types of recombination.

Despite its advantages, the mathematical complexity of the SLT model, described by a system of 7 differential equations, hinders its direct application for analyzing experimental data and estimating crucial physical parameters. Finding reliable analytical approximation is essential to simplify the model and enhance its practical utility.

The quasi-equilibrium (QE) approximation, also referred to as the quasi-steady assumption, has been widely applied in the theoretical analysis of TL and OSL phenomena, offering a pathway to simplify complex kinetic systems. A notable early application of this approach was proposed by Halperin and Braner (Halperin and Braner, 1960) who have suggested an approximate expression for the dependence of the emitted light on the relevant occupancies of traps and recombination centers. They made the simplifying assumptions, later called the quasi-equilibrium or quasi-steady assumptions $\left|\frac{dn_c}{dt}\right| \ll \left|\frac{dn}{dt}\right|, \left|\frac{dm}{dt}\right|$, which state that the rate of change of free carriers is significantly smaller than that of the trapped carriers, and that the concentration of free carriers is significantly smaller than that of trapped carriers $n_c \ll n, m$.

Such assumptions enabled Halperin and Braner to formulate an analytical expression for the emitted light in cases where charge carriers travel through the conduction band during read-out. The successful application, recognized limitations, and validity conditions of QE approximations in STM and LT models highlight their potential to simplify more complex models like SLT. This simplification can make SLT more tractable for experimental data analysis and parameter estimation, provided the applicability of QE is carefully validated.

Maxia (Maxia, 1978, 1980) examined the concept of QE by considering TL emission to be the result of two non-equilibrium phase transformation – namely, the release of an electron into the conduction band, and the recombination of the freed electron with a trapped hole. By assuming a linear relationship between the rate of entropy production and the transformation velocities for each of these reactions, Maxia applied Gibbs principle of minimum entropy production to arrive at the conclusion that the free carrier concentration does indeed remain approximately steady throughout the TL process. However, as pointed out by Chen and McKeever (Chen and Mckeever, 1997), the minimum entropy production principle is, in fact, just another way of imposing the limitation of quasi-steady-state on the free carrier concentrations, and the question whether the principle is valid during the TL process remains.

Chen et al. (Chen et al., 2012) and Pagonis et al. (Pagonis et al., 2013) when dealing with the semi localized transition (SLT) obtained analytical expressions for the TL intensity by assuming that excited states relax quite rapidly compared to the time scales of TL experiments. Chen assumed the excited state as quasi-steady.

This work investigates the applicability and validity of a QE approximation within the SLT framework. It is focused on the conditions under which QE assumptions hold true. Its validity is verified by solving the full set of SLT kinetic equations using specialized algorithms suitable for stiff differential systems, such as Radau methods implemented in Python. The goal is to derive and test simplified an analytical model based on QE conditions. We explore the behavior of the approximation under various physical conditions, including different trap parameters and heating rates, than, finally comparing their predictions against the full numerical solutions of the SLT model. The development of validated QE approximations for SLT would provide simpler, yet physically grounded, tools for interpreting TL and OSL experimental results, potentially leading to more accurate parameter estimation and a better understanding of luminescence kinetics in diverse materials.

## 2. Theory
### 2.1. The SLT model

To describe the kinetics of these processes analytically, the SLT model employs an enumeration technique. Instead of tracking global concentrations of trapped electrons and holes, the model defines variables representing the concentrations of T-RC pairs in different possible states of occupation. These states are typically denoted as $H_m^n$ (where the RC is full - an electron with a neighboring hole) and $E_m^n$ (where the RC is empty - an electron without a neighboring hole), with subscripts $m$ indicating the occupancy of the trap's ground states and superscripts $n$ indicating the occupancy of the trap's excited electron states. That states are a time dependent variable denoting concentration (in cm$^{-3}$).

The initial excitation generates only $H_1^0$ states, i.e. trapped hole–electron pairs. All other states are set to 0. The following diagram illustrates transitions between all states:

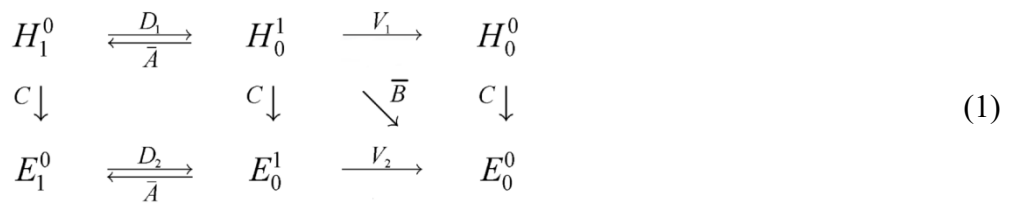 (1)

Key transitions incorporated into these equations include:

- ***D<sub>i</sub>* (Detrapping):** Thermal excitation of an electron from the trap's ground state to its local excited state. This is a thermally activated process.

$$D_1(t) = v_1 \exp\left(\frac{-E_1}{kT(t)}\right) \quad (2)$$

$$D_2(t) = v_2 \exp\left(\frac{-E_2}{kT(t)}\right) \quad (3)$$

- **$\bar{A}$ (Localized Trapping):** Transition of an electron from the local excited state back to the trap's ground state within the same T-RC pair.
- **$\bar{B}$ (Localized Recombination):** Recombination of an electron from the local excited state with the hole in the adjacent RC within the same T-RC pair.
- ***V<sub>i</sub>* (Delocalization/Excitation to Band):** Thermal excitation of an electron from the local excited state to the conduction band. This is also a thermally activated process.

$$V_1(t) = v_{v1} \exp\left(\frac{-E_{v1}}{kT(t)}\right) \quad (4)$$

$$V_2(t) = v_{v2} \exp\left(\frac{-E_{v2}}{kT(t)}\right) \quad (5)$$

- ***C* (Delocalized Recombination from the conduction band):** Recombination of a free electron from the conduction band with a hole in a RC.

In the first paper introducing SLT model (Mandowski, 2005) the coefficients were defined for both $H_m^n$ and $E_m^n$ states together. Nonetheless, as the states $H_m^n$ and $E_m^n$ are virtually different (i.e. an electron with and without a neighbouring hole, respectively), it is reasonable to allow different values of the coefficients *D* and *V* for these two classes (Mandowski, 2006). Therefore, we denote the coefficients as $D_1$(with $E_1$, $v_1$), $V_1$(with $E_{V1}$, $v_{v1}$), and $D_2$(with $E_2$, $v_2$), $V_2$(with $E_{V2}$, $v_{v2}$). For some physical reasons we may consider a "lonely electron" $E_m^n$ less bonded to the trap as compared to the electron in a hole–electron pair $H_m^n$. Therefore we expect $E_2 \leq E_1$ and $E_{V2} \leq E_{V1}$. These inequalities lead to the CD (cascade detrapping) phenomenon (Mandowski, 2006).

Using the diagram, we can derive a set of differential equations for these processes.

$$\dot{H}_1^0 = -(D_1 + Cn_c)H_1^0 + \bar{A}H_0^1 \quad (6a)$$

$$\dot{H}_0^1 = D_1 H_1^0 - (\bar{A} + \bar{B} + V_1 + Cn_c)H_0^1 \quad (6b)$$

$$\dot{H}_0^0 = V_1 H_0^1 - Cn_c H_0^0 \quad (6c)$$

$$\dot{E}_1^0 = Cn_c H_1^0 - D_2 E_1^0 + \bar{A} E_0^1 \quad (6d)$$

$$\dot{E}_0^1 = Cn_c H_0^1 + D_2 E_1^0 - (\bar{A} + V_2)E_0^1 \quad (6e)$$

$$\dot{E}_0^0 = \bar{B} H_0^1 + Cn_c H_0^0 + V_2 E_0^1 \quad (6f)$$

$$\dot{n}_c = -Cn_c(H_1^0 + H_0^1 + H_0^0) + V_1 H_0^1 + V_2 E_0^1 \tag{6g}$$

Luminescence intensity in SLT can arise from both localized recombination ($\mathcal{L}_{\bar{B}}$) and delocalized recombination ($\mathcal{L}_C$).

$$\mathcal{L}_{\bar{B}} = \bar{B} H_0^1 \tag{7}$$
$$\mathcal{L}_{\bar{B}} = Cn_c(H_1^0 + H_0^1 + H_0^0) \tag{8}$$

*2.2. Quasi-equilibrium conditions and the simplified equations*

To simplify the set of equations (6) we will assume that the two "intermediate" states in the diagram (1), i.e. $H_0^1$ and $E_0^1$ do not vary too much with respect to other variables, i.e. $\dot{H}_0^1 \approx 0$ and $\dot{E}_0^1 \approx 0$. The other, quite obvious assumption relates to the concentration of free electrons in the conduction band. The concentration is usually very small, i.e. $n_c \approx 0$. Therefore, considering equations (6b) and (6e) we come to the following approximate relations:

$$H_0^1 = H_1^0 \cdot \frac{D_1}{(\bar{A}+\bar{B}+V_1)} \tag{9}$$

$$E_0^1 = E_1^0 \cdot \frac{D_2}{(\bar{A}+V_2)} \tag{10}$$

These two variables can be eliminated from the set of eqs. (6) giving finally the reduced set of 5 differential equations:

$$\dot{H}_1^0 = -(D_1 + Cn_c) \cdot H_1^0 + \frac{\bar{A} \cdot D_1}{(\bar{A}+\bar{B}+V_1)} \cdot H_1^0 \tag{11a}$$

$$\dot{H}_0^0 = \frac{V_1 \cdot D_1}{(\bar{A}+\bar{B}+V_1)} \cdot H_1^0 - Cn_c H_0^0 \tag{11b}$$

$$\dot{E}_1^0 = Cn_c H_1^0 - D_2 E_1^0 + \frac{\bar{A} \cdot D_2}{(\bar{A}+V_2)} \cdot E_1^0 \tag{11c}$$

$$\dot{E}_0^0 = \frac{\bar{B} \cdot D_1}{(\bar{A}+\bar{B}+V_1)} \cdot H_1^0 + Cn_c H_0^0 + \frac{V_2 \cdot D_2}{(\bar{A}+V_2)} \cdot E_1^0 \tag{11d}$$

$$\dot{n}_c = -Cn_c \left( H_1^0 + H_0^0 + \frac{D_1}{(\bar{A}+\bar{B}+V_1)} \cdot H_1^0 \right) + \frac{V_1 \cdot D_1}{(\bar{A}+\bar{B}+V_1)} \cdot H_1^0 + \cdot \frac{V_2 \cdot D_2}{(\bar{A}+V_2)} \cdot E_1^0 \tag{11e}$$

We will call this the SLT-QE5 approximation. Luminescence intensities $\mathcal{L}_{\bar{B}}$ and $\mathcal{L}_C$ are defined by the same equations as in the previous case (7,8).

### 3. Numerical results

The set of approximate equations (11) SLT-QE5 was solved for various parameters and compared to the solutions of the genuine SLT model (6). The numerical solutions were calculated using the *solve_ivp* function from the *SciPy* library in *Python*. The *Radau* integration technique, chosen for its effectiveness in handling stiff differential equations, was employed to ensure accurate and stable results. Error control in numerical integration is managed by the *rtol* and *atol* parameters in the *solve_ivp* function.

Fig. 1 presents all variables and luminescence intensities calculated for SLT (6) and SLT-QE5 (11) models for the following parameters: $E_1 = 0.9\ eV$, $E_{V1} = 0.7\ eV$, $\nu_1 = \nu_2 = \nu_{\nu 1} = \nu_{\nu 2} = 10^{10} s^{-1}$ and the heating rate $\beta = 1\ Ks^{-1}$. The recombination coefficient $r = \bar{A}/\bar{B} = 1$. The simplest case was considered here with no change of activation energies during recombination. The agreement between the two models is quite good in the range where the TL peaks are visible.

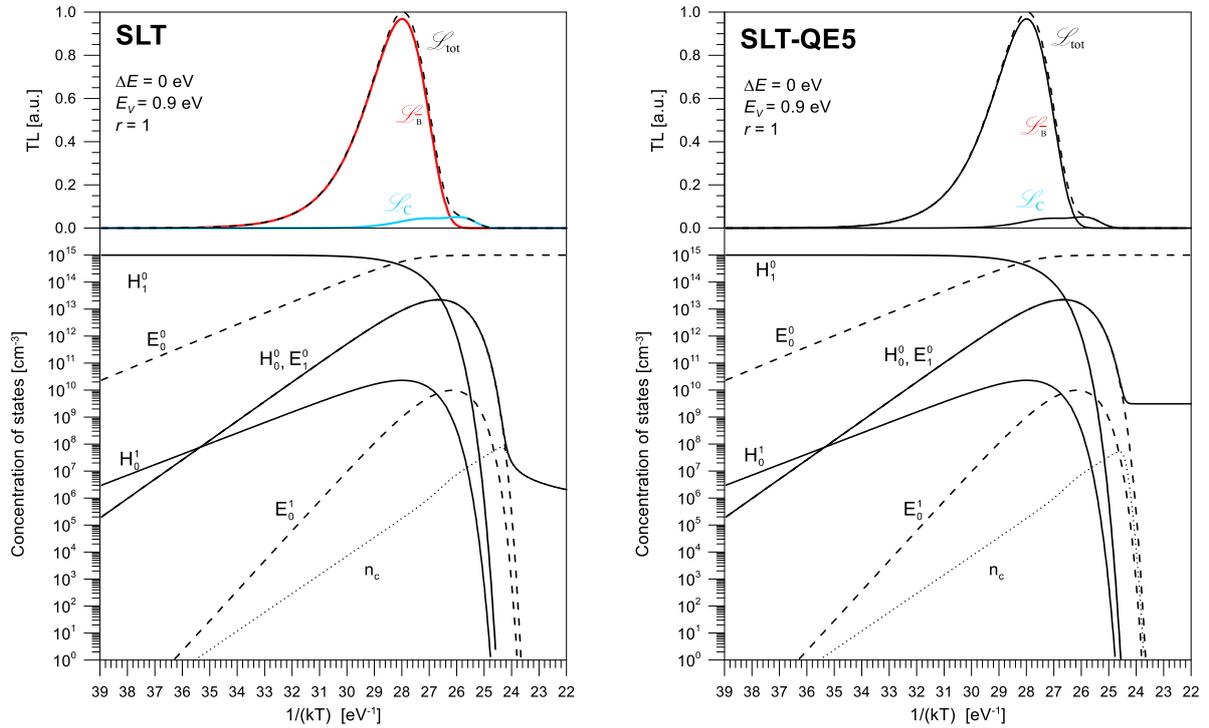

Fig. 1. Variables and luminescence intensities calculated for SLT (6) and SLT-QE5 (11) models for the following parameters: $E_1 = 0.9\ eV$, $E_{V1} = 0.7\ eV$, $\nu_1 = \nu_2 = \nu_{\nu 1} = \nu_{\nu 2} = 10^{10} s^{-1}$, heating rate $\beta = 1\ Ks^{-1}$ and the recombination coefficient $r = \bar{A}/\bar{B} = 1$.

The influence of the retrapping coefficient $r = \bar{A}/\bar{B}$ on the accuracy of the SLT-QE5 approximation is shown in Fig. 2. The relative deviation $\varepsilon$ of the total TL intensity $\mathcal{L}_{tot} = \mathcal{L}_{\bar{B}} + \mathcal{L}_C$ was calculated for $r$ ranging from $10^{-2}$ to $10^2$. The ε is defined as:

$$\varepsilon = \frac{\mathcal{L}_{tot.SLT} - \mathcal{L}_{tot.QE5}}{\mathcal{L}_{max.SLT}} \tag{12}$$

The accuracy decreases with decreasing retrapping coefficient. The largest values of the deviation (12) occur around the TL peak. However, they are small and even for $r = 10^{-2}$ do not exceed 0.013%. It should be noted that the calculations were performed for constant values of activation energy.

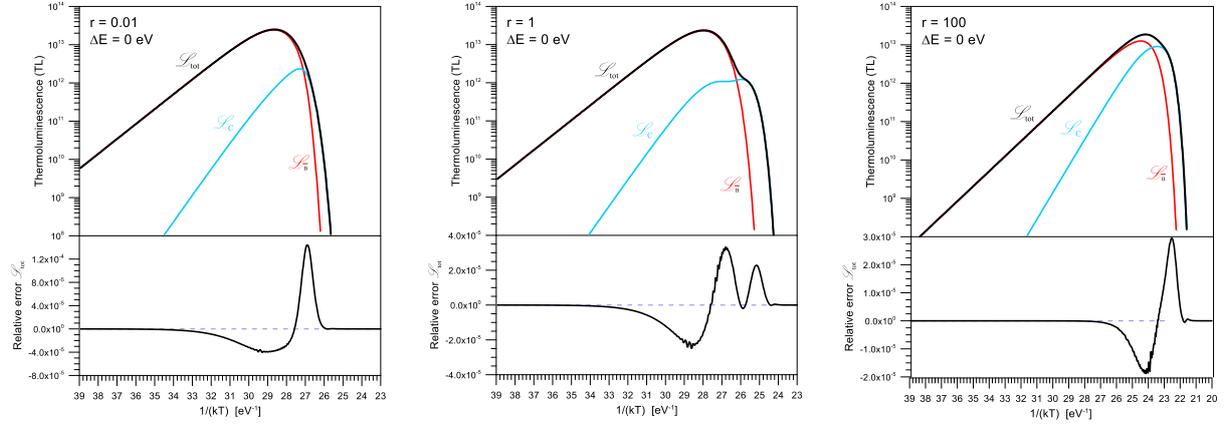

Fig. 2. Thermoluminescence glow curves calculated for SLT-QE5 (11) model for various retrapping coefficients. Bottom diagrams show relative deviation from the exact SLT model (6). Other parameters are the same as for the Fig. 1.

The other case with variable energy is shown in Fig. 3. For the calculation the value of $\Delta E = E_1 - E_2 = 0.2$ eV was assumed. This is the case where the *cascade detrapping* may occur producing extremely high frequency factors (Mandowski, 2006). In this case, the accuracy of SLT-QE5 approximation also increases with increasing value of the retrapping coefficient, but generally the accuracy is much lower. For $r = 10^{-2}$ $\varepsilon$ may be as high as 28% near the peak maximum. For $r = 1$ and $r = 10^2$ the maximal value decreases to 1.6% and 0.2%, respectively.

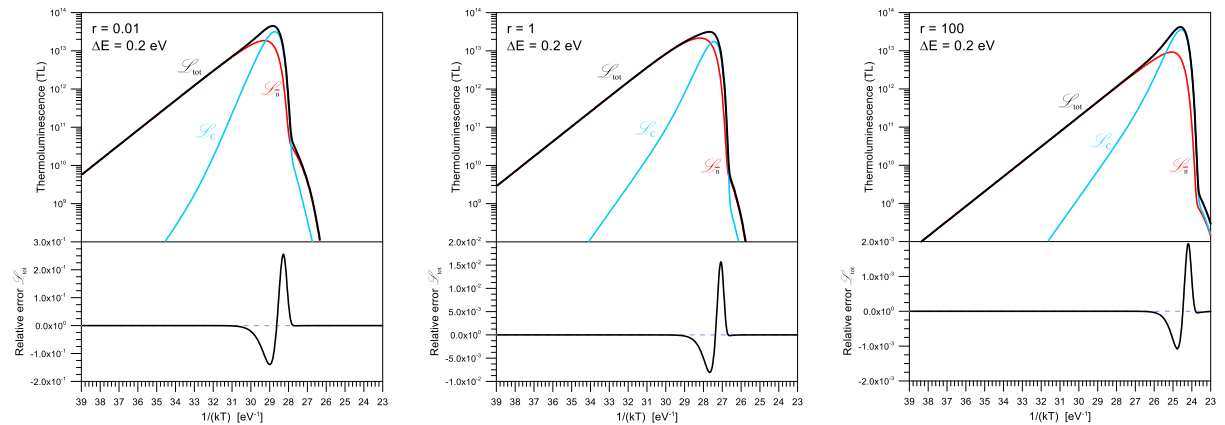

Fig. 3. Thermoluminescence glow curves calculated for SLT-QE5 (11) model for various retrapping coefficients in the case of variable detrapping activation energy D. The decrease of the energy $\Delta E = E_1 - E_2 = 0.2$ eV. Bottom diagrams show relative deviation from the exact SLT model (6). Other parameters are the same as for the Fig. 1.

## 4. Conclusions

The SLT model offers many advantages in describing complex TL curves measured in solid state detectors of ionizing radiation. The main drawback of this model is its mathematical complexity. This paper presents SLT-QE5 approximation, which reduces the number of differential equations to five. The accuracy of this approximation is very good except the case with large reduction of activation energy and very low retrapping. This probably leads to a significant accumulation of charge carriers in excited levels, resulting in a strong non-equilibrium state.

## Acknowledgements

This work was partly supported by research project no. 2018/31/B/ST10/03966 from the Polish National Science Centre.